\documentclass[10pt,conference]{IEEEtran}

\usepackage{amsmath}
\usepackage{bm}
\usepackage[ruled,vlined]{algorithm2e}
\usepackage[T1]{fontenc}
\usepackage{graphicx}
\usepackage[breaklinks]{hyperref}
\usepackage{breakurl}
\usepackage[utf8]{inputenc}
\usepackage{fancyvrb}
\usepackage{xcolor}
\usepackage{colortbl}
\usepackage{array}
\usepackage{subcaption}
\usepackage{multirow}
\usepackage{enumitem}
\usepackage{amssymb}
\usepackage{wrapfig}
\usepackage{pifont}% http://ctan.org/pkg/pifont
\usepackage{pgf-pie}
\usepackage{etoolbox}

\fvset{commandchars=\\\{\}}

\newcommand{\XComment}[1]{}
\newcommand{\XSpace}[1]{}
\newcommand{\CodeIn}[1]{\begin{small}\texttt{#1}\end{small}}

\newcommand{\DefMacro}[2]{\expandafter\newcommand\csname rmk-#1\endcsname{#2}}
\newcommand{\UseMacro}[1]{\csname rmk-#1\endcsname}
\newcommand{\TTitle}[1]{\textbf{#1}}

\definecolor{LightRed}{rgb}{1.0, 0.8, 0.8}
\definecolor{LightYellow}{rgb}{1.0, 1.0, 0.8}
\definecolor{LightGreen}{rgb}{0.8, 1.0, 0.8}
\definecolor{Gray}{gray}{0.8}
\definecolor{Green}{rgb}{0.0, 0.56, 0.0}
\definecolor{DarkRed}{rgb}{0.8, 0, 0}

\newcolumntype{L}[1]{m{#1}}
\newcolumntype{C}[1]{>{\centering\arraybackslash}m{#1}}
\newcolumntype{R}[1]{>{\raggedleft\arraybackslash}p{#1}}
\newcolumntype{N}{@{}m{0pt}@{}}

\newcommand{\Company}{Google}

\definecolor{cobalt}{rgb}{0.0, 0.28, 0.67}

\newcommand{\Var}[1]{\textit{#1}}

\DefMacro{ExploredASTNodesBeforeFindingFault}{9}
\DefMacro{ExploredASTNodesAtLeastFindAFault}{10}
\DefMacro{ASTNodesInFigure}{19}
\DefMacro{ASTNodesInModel}{169}
\DefMacro{ExploredASTNodesPercentage}{5.9}
\DefMacro{NotInspectedASTNodesPercentage}{94.1}

\DefMacro{qps-avg-min}{6}
\DefMacro{qps-avg-max}{27}
\DefMacro{qps-min-min}{0}
\DefMacro{qps-min-max}{12}
\DefMacro{qps-median-min}{4}
\DefMacro{qps-median-max}{24}
\DefMacro{qps-max-min}{18}
\DefMacro{qps-max-max}{89}
\DefMacro{latency-avg-min}{265}
\DefMacro{latency-avg-max}{414}
\DefMacro{latency-min-min}{1}
\DefMacro{latency-min-max}{81}
\DefMacro{latency-median-min}{150}
\DefMacro{latency-median-max}{179}
\DefMacro{latency-max-min}{69k}
\DefMacro{latency-max-max}{886k}
\DefMacro{stream-target-count-avg}{674}
\DefMacro{stream-target-count-min}{1}
\DefMacro{stream-target-count-median}{2}
\DefMacro{stream-target-count-max}{70m}
\DefMacro{stream-build-count-avg}{2}
\DefMacro{stream-build-count-min}{1}
\DefMacro{stream-build-count-median}{1}
\DefMacro{stream-build-count-max}{84k}
\DefMacro{build-target-count-avg}{339}
\DefMacro{build-target-count-min}{1}
\DefMacro{build-target-count-median}{24}
\DefMacro{build-target-count-max}{900}
\DefMacro{build-execution-time-avg}{394}
\DefMacro{build-execution-time-min}{0.1}
\DefMacro{build-execution-time-median}{250}
\DefMacro{build-execution-time-max}{7923}
\DefMacro{build-memory-avg}{3.87}
\DefMacro{build-memory-min}{0.04}
\DefMacro{build-memory-median}{2.82}
\DefMacro{build-memory-max}{13.26}
\DefMacro{build-occupancy-avg}{18.3}
\DefMacro{build-occupancy-min}{0}
\DefMacro{build-occupancy-median}{2.8}
\DefMacro{build-occupancy-max}{2066.4}

\DefMacro{only-one-target-build-cnt}{21.6m}
\DefMacro{only-one-target-build-percentage}{21.12}
\DefMacro{only-one-target-oom-cnt}{22.5k}
\DefMacro{only-one-target-oom-percentage}{28.13}
\DefMacro{only-one-target-oom-cnt-ratio}{0.10}
\DefMacro{only-one-target-de-cnt}{6.6k}
\DefMacro{only-one-target-de-percentage}{14.22}
\DefMacro{only-one-target-de-cnt-ratio}{0.03}
\DefMacro{only-one-target-avg-target-cnt}{1}
\DefMacro{only-one-target-avg-execution-time}{241}
\DefMacro{only-one-target-avg-memory}{2.2}
\DefMacro{only-one-target-avg-occupancy}{2.0}
\DefMacro{only-one-target-median-target-cnt}{1}
\DefMacro{only-one-target-median-execution-time}{92}
\DefMacro{only-one-target-median-memory}{1.6}
\DefMacro{only-one-target-median-occupancy}{0.3}
\DefMacro{max-targets-build-cnt}{33.7m}
\DefMacro{max-targets-build-percentage}{32.95}
\DefMacro{max-targets-oom-cnt}{1.9k}
\DefMacro{max-targets-oom-percentage}{2.34}
\DefMacro{max-targets-oom-cnt-ratio}{0.01}
\DefMacro{max-targets-de-cnt}{6.8k}
\DefMacro{max-targets-de-percentage}{14.59}
\DefMacro{max-targets-de-cnt-ratio}{0.02}
\DefMacro{max-targets-avg-target-cnt}{900}
\DefMacro{max-targets-avg-execution-time}{371}
\DefMacro{max-targets-avg-memory}{4.0}
\DefMacro{max-targets-avg-occupancy}{33.4}
\DefMacro{max-targets-median-target-cnt}{900}
\DefMacro{max-targets-median-execution-time}{275}
\DefMacro{max-targets-median-memory}{3.7}
\DefMacro{max-targets-median-occupancy}{18.8}
\DefMacro{all-remaining-targets-build-cnt}{33.8m}
\DefMacro{all-remaining-targets-build-percentage}{33.06}
\DefMacro{all-remaining-targets-oom-cnt}{7.6k}
\DefMacro{all-remaining-targets-oom-percentage}{9.52}
\DefMacro{all-remaining-targets-oom-cnt-ratio}{0.02}
\DefMacro{all-remaining-targets-de-cnt}{21.0k}
\DefMacro{all-remaining-targets-de-percentage}{45.17}
\DefMacro{all-remaining-targets-de-cnt-ratio}{0.06}
\DefMacro{all-remaining-targets-avg-target-cnt}{67}
\DefMacro{all-remaining-targets-avg-execution-time}{403}
\DefMacro{all-remaining-targets-avg-memory}{3.2}
\DefMacro{all-remaining-targets-avg-occupancy}{13.2}
\DefMacro{all-remaining-targets-median-target-cnt}{5}
\DefMacro{all-remaining-targets-median-execution-time}{227}
\DefMacro{all-remaining-targets-median-memory}{2.4}
\DefMacro{all-remaining-targets-median-occupancy}{2.5}
\DefMacro{max-memory-build-cnt}{13.0m}
\DefMacro{max-memory-build-percentage}{12.71}
\DefMacro{max-memory-oom-cnt}{46.9k}
\DefMacro{max-memory-oom-percentage}{58.75}
\DefMacro{max-memory-oom-cnt-ratio}{0.36}
\DefMacro{max-memory-de-cnt}{11.2k}
\DefMacro{max-memory-de-percentage}{24.13}
\DefMacro{max-memory-de-cnt-ratio}{0.09}
\DefMacro{max-memory-avg-target-cnt}{154}
\DefMacro{max-memory-avg-execution-time}{679}
\DefMacro{max-memory-avg-memory}{8.1}
\DefMacro{max-memory-avg-occupancy}{15.2}
\DefMacro{max-memory-median-target-cnt}{9}
\DefMacro{max-memory-median-execution-time}{574}
\DefMacro{max-memory-median-memory}{8.3}
\DefMacro{max-memory-median-occupancy}{2.6}
\DefMacro{max-occupancy-build-cnt}{59.3k}
\DefMacro{max-occupancy-build-percentage}{0.06}
\DefMacro{max-occupancy-oom-cnt}{35}
\DefMacro{max-occupancy-oom-percentage}{0.04}
\DefMacro{max-occupancy-oom-cnt-ratio}{0.06}
\DefMacro{max-occupancy-de-cnt}{855}
\DefMacro{max-occupancy-de-percentage}{1.84}
\DefMacro{max-occupancy-de-cnt-ratio}{1.4}
\DefMacro{max-occupancy-avg-target-cnt}{506}
\DefMacro{max-occupancy-avg-execution-time}{1424}
\DefMacro{max-occupancy-avg-memory}{5.5}
\DefMacro{max-occupancy-avg-occupancy}{471.9}
\DefMacro{max-occupancy-median-target-cnt}{555}
\DefMacro{max-occupancy-median-execution-time}{1301}
\DefMacro{max-occupancy-median-memory}{5.6}
\DefMacro{max-occupancy-median-occupancy}{521.6}
\DefMacro{memory-estimate-error-build-cnt}{104.3k}
\DefMacro{memory-estimate-error-build-percentage}{0.10}
\DefMacro{memory-estimate-error-oom-cnt}{967}
\DefMacro{memory-estimate-error-oom-percentage}{1.21}
\DefMacro{memory-estimate-error-oom-cnt-ratio}{0.93}
\DefMacro{memory-estimate-error-de-cnt}{25}
\DefMacro{memory-estimate-error-de-percentage}{0.05}
\DefMacro{memory-estimate-error-de-cnt-ratio}{0.02}
\DefMacro{memory-estimate-error-avg-target-cnt}{239}
\DefMacro{memory-estimate-error-avg-execution-time}{443}
\DefMacro{memory-estimate-error-avg-memory}{3.0}
\DefMacro{memory-estimate-error-avg-occupancy}{21.0}
\DefMacro{memory-estimate-error-median-target-cnt}{300}
\DefMacro{memory-estimate-error-median-execution-time}{306}
\DefMacro{memory-estimate-error-median-memory}{1.8}
\DefMacro{memory-estimate-error-median-occupancy}{11.9}
\DefMacro{occupancy-estimate-error-build-cnt}{716}
\DefMacro{occupancy-estimate-error-build-percentage}{0.00}
\DefMacro{occupancy-estimate-error-oom-cnt}{0}
\DefMacro{occupancy-estimate-error-oom-percentage}{0.00}
\DefMacro{occupancy-estimate-error-oom-cnt-ratio}{0.00}
\DefMacro{occupancy-estimate-error-de-cnt}{0}
\DefMacro{occupancy-estimate-error-de-percentage}{0.00}
\DefMacro{occupancy-estimate-error-de-cnt-ratio}{0.00}
\DefMacro{occupancy-estimate-error-avg-target-cnt}{222}
\DefMacro{occupancy-estimate-error-avg-execution-time}{391}
\DefMacro{occupancy-estimate-error-avg-memory}{2.2}
\DefMacro{occupancy-estimate-error-avg-occupancy}{19.6}
\DefMacro{occupancy-estimate-error-median-target-cnt}{300}
\DefMacro{occupancy-estimate-error-median-execution-time}{273}
\DefMacro{occupancy-estimate-error-median-memory}{1.6}
\DefMacro{occupancy-estimate-error-median-occupancy}{11.3}
\DefMacro{sum-build-cnt}{102.3m}
\DefMacro{sum-oom-cnt}{79.9k}
\DefMacro{sum-oom-cnt-ratio}{0.08}
\DefMacro{sum-de-cnt}{46.5k}
\DefMacro{sum-de-cnt-ratio}{0.05}
\DefMacro{type-i-de-percentage}{1.32}
\DefMacro{type-ii-de-percentage}{98.68}

\begin{document}

\title{Smart Build Targets Batching Service at \Company}

%% \author{Kaiyuan Wang, Allison Sullivan$^{\dagger}$, Darko Marinov*, and Sarfraz Khurshid}
%% \affiliation{%
%%   \institution{The University of Texas at Austin, $^{\dagger}$North Carolina A\&T State University, *University of Illinois at Urbana-Champaign}
%% }
%% \email{{kaiyuanw, khurshid}@utexas.edu, allisonksullivan@gmail.com, marinov@illinois.edu}

% \renewcommand{\shortauthors}{K. Wang et al.}

\makeatletter
\newcommand{\linebreakand}{%
  \end{@IEEEauthorhalign}
  \hfill\mbox{}\par
  \mbox{}\hfill\begin{@IEEEauthorhalign}
}
\makeatother

\author{\IEEEauthorblockN{Kaiyuan Wang}
\IEEEauthorblockA{
\textit{Google Inc} \\
%% Sunnyvale, CA USA \\
kaiyuanw@google.com}
\and
\IEEEauthorblockN{Daniel Rall}
\IEEEauthorblockA{
\textit{Google Inc} \\
%% Sunnyvale, CA USA \\
dlr@google.com}
\and
\IEEEauthorblockN{Greg Tener}
\IEEEauthorblockA{
\textit{Google Inc} \\
%% Sunnyvale, CA USA \\
gtener@google.com}
\linebreakand
\IEEEauthorblockN{Vijay Gullapalli}
\IEEEauthorblockA{
\textit{Google Inc} \\
%% Sunnyvale, CA USA \\
vijaysagar@google.com}
\and
\IEEEauthorblockN{Xin Huang}
\IEEEauthorblockA{
\textit{Google Inc} \\
%% Sunnyvale, CA USA \\
xnh@google.com}
\and
\IEEEauthorblockN{Ahmed Gad}
\IEEEauthorblockA{
\textit{Google Inc} \\
%% Sunnyvale, CA USA \\
ahmedgad@google.com}
}

\makeatletter
\patchcmd{\@maketitle}
  {\addvspace{0.5\baselineskip}\egroup}
  {\addvspace{-1\baselineskip}\egroup}
  {}
  {}
\makeatother

\maketitle

\begin{abstract}
\Company{} has a monolithic codebase with tens of millions build
targets.  Each build target specifies the information that is needed
to build a software artifact or run tests.  It is common to execute a
subset of build targets at each revision and make sure that the change
does not break the codebase.  \Company's build service system uses
Bazel to build targets.  Bazel takes as input a build that specifies
the execution context, flags and build targets to run.  The outputs
are the build libraries, binaries or test results.  To be able to
support developer's daily activities, the build service system runs
millions of builds per day.

It is a known issue that a build with many targets could run out of
the allocated memory or exceed its execution deadline.  This is
problematic because it reduces the developer's productivity, e.g. code
submissions or binary releases.  In this paper, we propose a technique
that predicts the memory usage and executor occupancy of a build.  The
technique batches a set of targets such that the build created with
those targets does not run out of memory or exceed its deadline.  This
approach significantly reduces the number of builds that run out of
memory or exceed the deadlines, hence improving developer's
productivity.
\end{abstract}

%% \begin{IEEEkeywords}
%% Alloy, Fault localization, AlloyFL
%% \end{IEEEkeywords}

\section{Introduction}
\label{sec:introduction}

\Company{} has a monolithic codebase and it grows rapidly with an
average of more than 100,000 code submissions per day.  To make sure
that the new changes do not break the existing codebase, \Company{}
has adopted Continuous Integration
(CI)~\cite{duvall2007continuous,memon2017taming}.  For each code
change, a CI system first uses the build service
system~\cite{scalable-build-service} to build the affected libraries
and binaries.  Then, it runs the affected tests to check if the built
libraries and binaries are working as intended.

\Company's build service system uses Bazel~\cite{bazel} to build
libraries and run tests.  Bazel takes as input a set of build
specification files that declare build targets.  We refer to a build
target as a target in the rest of this paper.  A target specifies what
is needed to produce an artifact, such as a library or binary.  A test
target specifies what is needed to run tests and check if a code
change breaks the codebase.  Bazel decides how to build a given target
based on the target's specification.

As the codebase grows, each change may affect a lot of targets.  For
example, a code change on a common library like Guava~\cite{guava}
could affect many Java targets.  A C++ compiler change could affect
all C++ targets.  Moreover, the postsubmit service needs to guarantee
that all targets in a given change revision compile and pass all
tests.  The above use cases are common and require Bazel to build up
to millions of targets at once.  Running a large number of targets in
a build execution may cause \emph{out of memory} (\emph{OOM}) errors
in Bazel as it exceeds the memory of a single machine for dependency
analysis.  Moreover, \Company's infrastructure limits the number of
executors, e.g. CPU, GPU and TPU, a build can use in parallel at the
build execution time.  So executing too many targets in a build may
cause the build to take too long and exceed its invocation deadline,
resulting in \emph{deadline exceeded} (\emph{DE}) errors.  In
contrary, executing too few targets in a build may use up all the
machines allocated to run builds.

In this paper, we propose a build target batching service
(\emph{BTBS}) that partitions a large stream of targets into batches
and creates a build for each batch of targets such that those builds
do not have OOM or DE errors.  The technique relies on a memory
estimation model and an executor occupancy estimation model.  The
memory model predicts the peak memory usage of a build.  The occupancy
model predicts the average executor occupancy of a build.  The
technique partitions a large stream of targets into target batches
such that the build with the flags and each batch of targets consumes
a limited memory or executor occupancy.  The results show that BTBS
generates few OOM and DE builds with \UseMacro{sum-oom-cnt-ratio}\%
OOM rate and \UseMacro{sum-de-cnt-ratio}\% DE rate, which saves a lot
of computational resources used for build failure retries.
\XComment{We believe that the technique is also beneficial for other
  users who use build tools that operate on targets,
  e.g. Bazel~\cite{bazel} or Buck~\cite{buck}, and face a similar
  problem.}

The paper makes the following contributions:

\begin{itemize}[leftmargin=*]

\item It presents the first technique that effectively creates builds
  from a large stream of targets with the goal of minimizing the
  number of OOM and DE errors.  It is also the first technique that
  predicts memory and occupancy usages of build executions.

\item It demonstrates that the technique is able to reduce the OOM
  rate to \UseMacro{sum-oom-cnt-ratio}\% and the DE rate to
  \UseMacro{sum-de-cnt-ratio}\%.  Our past experience shows that BTBS
  is critical and improves the developer's productivity by reducing
  build failure retries due to OOM or DE errors.
\end{itemize}

\section{Background}
\label{sec:background}

\subsection{Bazel}
\label{sec:sec:bazel}

{
\setlength{\extrarowheight}{1.5pt}
\begin{table*}[!t]
  \centering
  \caption{Example build flags}
  \label{tab:build-flag-examples}
  \normalsize
  \begin{tabular}{ >{\centering\arraybackslash}m{3cm} @{\hspace{6ex}}>{\raggedright\arraybackslash}m{14cm} } 
  \hline
  \TTitle{Category} & \multicolumn{1}{c}{\TTitle{Example}} \\
  \hline
  %% Package location & --package\_path specifies the searching directories to find the build specification file. \\
  Error checking & --check\_visibility enables checking if all dependent targets are visible to all targets to build. \\
  Tool flags & --copt specifies the arguments to be passed to the C++ compiler. \\
  Build semantics & --cpu specifies the target CPU architecture to be used for the compilation. \\
  %% Version & --compiler specifies the C++ compiler version. \\
  Execution strategy & --jobs specifies a limit on the number of concurrently running executors during build execution. \\
  Output selection & --fuseless\_output restricts Bazel to generate intermediate output files in memory. \\
  %% Output selection & \CodeIn{--test\_filter} specifies a filter that the test runner uses to pick a subset of test for running. \\
  %% Verbosity & --show\_progress causes build progress messages to be displayed. \\
  %% Workspace status & --stamp controls whether stamping is enabled for rule types that support it. \\
  Platform & --platforms specifies the labels of the platform rules describing the target platforms. \\
  Miscellaneous & --use\_action\_cache enables Bazel's local action cache. \\
  \hline
\end{tabular}

\end{table*}
}

\begin{figure*}[!t]
  \captionsetup[subfigure]{aboveskip=-7pt,belowskip=-2pt}
  \centering
  \normalsize
  \include{figures/build-target-examples}
  \caption{Example Java targets}
  \label{fig:build-target-examples}
\end{figure*}

Bazel is an open-source build and test tool and is widely used within
\Company.  It is responsible for transforming source code into
libraries, executable binaries, and other artifacts.  Bazel takes as
input a set of flags and targets that programmers declare in build
files.  It supports a large number of command line options and these
options can affect the way Bazel generates outputs.

Table~\ref{tab:build-flag-examples} shows different flag categories
with examples.  For example, the \CodeIn{--fuseless\_output} flag
restricts Bazel to generate intermediate output in memory instead of
writing it to the disk.  The \CodeIn{--jobs} flag specifies a limit on
the number of concurrently running executors during a build execution.
So \CodeIn{--fuseless\_output} and \CodeIn{--jobs} can significantly
affect the memory usage and executor occupancy of a build,
respectively.

Figure~\ref{fig:build-target-examples} shows some example Java
targets.  \CodeIn{HelloWorld} is a Java library target that compiles
\CodeIn{HelloWorld.java} into a library (\CodeIn{java\_library} is the
target rule).  \CodeIn{HelloWorldTest} is a Java library target that
compiles \CodeIn{HelloWorldTest.java} into a library and it depends on
the \CodeIn{HelloWorld} target because \CodeIn{HelloWorldTest.java}
uses \CodeIn{HelloWorld.java}.  \CodeIn{AllTests} is a Java test
target that runs the \CodeIn{HelloWorldTest} library using JUnit and
the execution requires GPU.  When a programmer issues a command to
build a target, Bazel first ensures that the required dependencies of
the target are built.  Then, it builds the desired target from its
sources and dependencies.  When a programmer issues a command to run a
test target, Bazel will first build all dependencies of the test
target and then execute the tests.

\subsection{Build Service System}
\label{sec:sec:build-service-system}

\begin{figure}[!t]
  \centering
  \includegraphics[width=.49\textwidth]{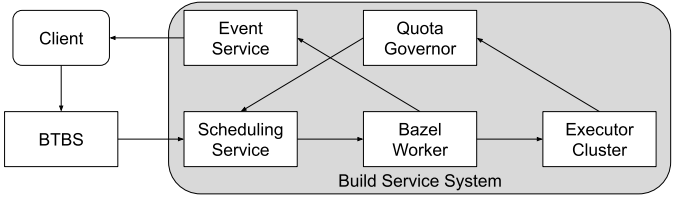}
  \caption{Build service architecture}
  \label{fig:build-service-architecture}
\end{figure}

BTBS takes as input a set of flags and a stream of targets, and
outputs a set of builds that include all targets with the same flags.
The output builds are sent to the build service
system~\cite{scalable-build-service} for execution.
Figure~\ref{fig:build-service-architecture} shows how BTBS is
connected to the build service system.  The system diagram is
simplified for brevity.

\XComment{The clients of BTBS typically compute the set of targets
  that need to be built or tested for a given task.  For example, the
  Test Automation Platform~\cite{memon2017taming} (TAP) has a
  presubmit service and a postsubmit service.  The presubmit service
  computes the set of affected targets for a given code change and try
  to make sure that the code change compiles and is well tested.  The
  postsubmit collect all targets after a set of code changes to
  guarantee that all targets can be built and executed successfully at
  a given revision.  Both services send the set of targets to BTBS and
  they could send up to millions of targets at once.  Another client
  of BTBS is the code coverage service and it could send millions of
  test targets to collect code coverage~\cite{codeoverageatgoogle}.
  The compiler testing service also sends millions of targets to check
  if a new compiler version breaks existing targets.  Building an
  efficient and effective target batching algorithm on the client side
  is tedious and wastes a lot of engineering effort.  So BTBS serves
  as a central point and provides a effective target batching service
  for all clients that need to build a large number of targets.  After
  all targets are sent to BTBS, the clients receive back a set of
  build request IDs that can be used to query the build status.}

BTBS splits a stream of targets into batches and creates a build for
each batch of targets.  Those builds are enqueued to the scheduling
service.  The scheduling service waits until there are available
resources and dequeues a build to a Bazel worker for execution.  The
worker allocates a fixed amount of memory for each Bazel process and a
build runs out of memory if the Bazel process uses more than the
allocated memory during execution.  Bazel translates the build flags
and targets into actions, and sends those actions to the executor
cluster for the actual execution.  For example, if a test requires GPU
then Bazel will send it to the GPU executors.  The executor cluster
has a large but limited number of executors and each executor talks to
a Bazel process to execute actions.  The executor cluster also has an
action cache to minimize duplicate work.  Each Bazel process is
configured to use a limited number of executors concurrently to avoid
the case where a very large build uses a lot of executors and blocks
other build executions.  As a consequence, executing a large build
that could use more executors than the limit causes the additional
actions to queue until some executors become idle again.  This may
cause the build to exceed its deadline and we call these builds
\CodeIn{Type I} DE builds.  It is also possible that some actions
depend on other actions and they must be executed in sequence.  This
may cause build to exceed its deadline if some actions are long
running.  We call these builds \CodeIn{Type II} DE builds.  An action
uses both memory and executor in the executor cluster.  We use an
executor service unit (ESU) to unify the expense of both
\emph{executor} memory and CPU, and 1 ESU is equal to 2.5GB of memory
or 1 executor.  Note that the executor memory usage is different from
the Bazel memory usage.  We refer to the occupancy usage as the ESU
used by a build in the rest of the paper.  BTBS reduces \CodeIn{Type
  I} DE errors by limiting the occupancy of a build, and we do not
consider \CodeIn{Type II} DE errors because it is not correlated with
the build occupancy usage.  BTBS assumes that a build with too much
ESU usage is more likely to have queuing actions that cause
\CodeIn{Type I} DE errors.  The executor cluster reports the executor
availability to the quota governor and the scheduling service makes
dequeuing decisions based on the executor availability from the quota
governor.  During a build execution, the Bazel process reports the
progress and result to the build event service.  Clients of BTBS can
use the build request IDs to query the build event service and find
the build progress and status in real time.

\subsection{Collateral Damage}
\label{sec:sec:collateral-damage}

Each build may use different executor types.  For example, the
\CodeIn{AllTests} Java test target can use both x86 CPU and GPU during
execution.  \XComment{Some targets need to be executed on iOS
  devices.}  The build scheduling service throttles builds based on
the availability of each executor type.  For example, a build that
requires both x86 CPU and GPU can only be dequeued when both x86 CPU
executors and GPU executors are available.  In other words, a build
might be delayed if one of its required executor types is unavailable
even if all other required executor types are available.  Moreover,
the delayed build will reserve the executor resources and block other
lower priority builds from dequeuing.  This design prevents high
priority builds from being starved by lower priority builds that use
very few ESU~\cite{scalable-build-service}.  As a consequence, a build
that requires more executor types is more likely to be delayed and it
may block other lower priority builds from dequeuing.

If a delayed build contains targets that require different sets of
executor types, then some of the targets could have been dequeued if
they are in a separate build.  For example, assume that a build
contains one target that only uses x86 executors and another target
that uses both x86 and GPU executors.  The build could be throttled
due to insufficient GPU executors but the target that only uses x86
executors could be dequeued if it is built separately.  We call such
delays \emph{collateral damage}.

\subsection{Linear Regression}
\label{sec:sec:linear-regression}

In machine learning, regression analysis is a set of statistical
processes for estimating the relationships between a dependent
variable (denoted as $y$) and one or more independent variables
(denoted as $x$).  The most common form of regression analysis is
linear regression~\cite{montgomery2012introduction}.  It tries to find
the line that most closely fits the data according to a specific
mathematical criterion.  For example, the method of ordinary least
squares computes the unique line (or hyperplane) that minimizes the
sum of squared distances between the true data and that line (or
hyperplane).

Given a set of data set $\{y_{i},x_{i1},\ldots,x_{ip}\}_{i=1}^{n}$ of
n statistical units.  A simple linear regression model has the
following form:
$$y_{i}=\beta_{0}+\beta_{1}x_{i1}+\cdots+\beta_{p}x_{ip}+\varepsilon_{i}=\mathbf{x}_{i}^{\mathsf{T}}{\boldsymbol{\beta}}+\varepsilon_{i},\qquad i=1,\ldots ,n$$

where $^{\mathsf{T}}$ denotes the transpose, so that
$\mathbf{x}_{i}^{\mathsf{T}}{\boldsymbol{\beta}}$ is the inner product
between vectors $\mathbf{x}_{i}$ and $\boldsymbol{\beta}$.

Fitting a linear model to a given data set usually requires estimating
the regression coefficients $\boldsymbol{\beta}$ such that the error
term
$\boldsymbol{\varepsilon_{i}}=\mathbf{y_i}-\mathbf{x}_{i}^{\mathsf{T}}{\boldsymbol{\beta}}$
is minimized across all $n$ samples.  For example, it is common to use
the sum of squared errors $\sum_{i=1}^{n}\varepsilon^2_{i}$ as the
quality of the fit.  Linear models can be efficiently trained using
stochastic gradient descent~\cite{bottou2010large}.  \XComment{There
  are many frameworks that support training linear regression models,
  e.g. Scikit learn~\cite{pedregosa2011scikit} and
  Tensorflow~\cite{abadi2016tensorflow}.  In this paper, we use the
  Sibyl~\cite{chandra2010sibyl} framework which is similar to
  Tensorflow.}

\subsection{Feature Cross}
\label{sec:sec:feature-cross}

A feature cross is a synthetic feature formed by multiplying
(crossing) two or more features~\cite{featurecross}. Crossing
combinations of features can provide predictive abilities beyond what
those features can provide individually.  As a consequence, feature
crosses can help us learn non-linear functions using linear
regression.  A well-known example is that the XOR function $f(x,y)$
where $x,y,f(x,y)\in\{0,1\}$ is not linearly separable and it cannot
be written as $f(x,y)=\alpha{x}+\beta{y}+\gamma$ where $\alpha$,
$\beta$ and $\gamma$ are real numbers.  However, the XOR function can
be written as $f(x,y)=x+y-2xy$, where the $xy$ term is a feature cross
for $x$ and $y$.

In practice, machine learning models seldom cross continuous features.
However, machine learning models do frequently cross one-hot feature
vectors.  Feature crosses of one-hot feature vectors are analogous to
logical conjunctions.  For example, suppose we bin latitude and
longitude, producing separate one-hot five-element feature vectors,
e.g. [0, 0, 0, 1, 0].  Further assume that we want to predict the
gross income of a person using the binned latitude and longitude as
features.  Creating a feature cross of these two features results in a
25-element one-hot vector (24 zeroes and 1 one).  The single 1 in the
cross identifies a particular conjunction of latitude and longitude.
By feature crossing these two one-hot vectors, the model can form
different conjunctions, which will ultimately produce far better
results compared to a linear combination of individual features.

\XComment{Linear learners scale well to massive data.  Using feature
  crosses on massive data sets is one efficient strategy for learning
  highly complex models.}

\section{Build Target Batching Service}
\label{sec:btbs}

\subsection{EnqueueTargets API}
\label{sec:sec:api}

BTBS provides a single remote procedure call (RPC) called
\CodeIn{EnqueueTargets}.  The RPC is a streaming RPC and it takes as
input a sequence of requests and returns to the clients a sequence of
responses.  The main reason to have a streaming RPC is to avoid
needing to stream millions of targets in a single request.  We enforce
the first \CodeIn{EnqueueTargets} request to include the execution
context, build flags and optionally some targets.  The execution
context points to either a workspace that contains the unsubmitted
code, or an existing code revision.  The flags and targets are
described in Section~\ref{sec:sec:bazel}.  The subsequent requests may
only contain the remaining targets.  BTBS will create a set of builds
with the same execution context and flags but different targets.  The
created builds include all targets in the requests.  Each
\CodeIn{EnqueueTargets} response contains the enqueued build request
ID so that the client can use it to query the build status.

\subsection{Group Targets}
\label{sec:sec:group-targets}

Once BTBS receives all targets from the client, it first groups
targets by the executor types they use.  This avoids the collateral
damage as described in Section~\ref{sec:sec:collateral-damage}.  For
example, all targets that use only x86 executors will be grouped
together.  All targets that use both x86 and Mac executors will be
grouped together.  Given a build target, BTBS determines its required
executor types by checking the \CodeIn{tags} attribute, e.g. the
\CodeIn{requires-gpu} tag in the \CodeIn{AllTests} target in
Figure~\ref{fig:build-target-examples}.  If a target does not have a
tag, then BTBS determines the required executor type by the target
rule.  For example, if a target has the rule \CodeIn{ios\_ui\_test},
then it uses Mac executors.

Once all targets are grouped by their required executor types, we sort
each group of targets in lexicographical order.  This is a heuristic
to increase the probability that the batched targets share similar
dependencies and thus allow Bazel to construct tighter dependency
graphs and reduce the memory usage.  The idea is that developers tend
to declare targets for the same project under the same directory.  The
targets in the same directory are likely to share dependent targets.
For example, target \CodeIn{//a/b/c:t1} may share more dependencies
with target \CodeIn{//a/b/c:t2} than target \CodeIn{//d/e/f:t3}.  The
assumption may not hold for all targets, but this heuristic works well
in practice.

\subsection{Batch Targets}
\label{sec:sec:batch-targets}

\begin{algorithm}[!t]
  \small
  \caption{Batching targets algorithm\label{alg:batching-targets}}
  \DontPrintSemicolon
% \Indm
\SetKw{Continue}{continue}
\SetKw{Break}{break}
\SetKw{true}{True}
\SetKw{false}{False}
\SetKwFunction{FSearch}{limitBatchSizeByCutoff}
\SetKwFunction{FEstimate}{getEstimateFromTargets}
\SetKwProg{Fn}{Function}{:}{}
\KwIn{List of targets to batch \Var{allTargets};
  Max targets per batch \Var{maxTargetsPerBatch};
  Memory cutoff value \Var{memoryCutoff};
  Occupancy cutoff value \Var{occupancyCutoff}.}
\KwOut{Target batches \Var{batches}.}

% \Indp
\BlankLine
\Var{batches} = [] \\
\While{len(\Var{allTargets}) > 0} {
  \Var{targets} = \Var{allTargets}[:\Var{maxTargetsPerBatch}] \\
  \Var{targets} = \FSearch{\Var{targets}, \Var{memoryCutoff}, \Var{MEMORY\_MODEL}} \\
  \Var{targets} = \FSearch{\Var{targets}, \Var{occupancyCutoff}, \Var{OCCUPANCY\_MODEL}} \\
  \Var{batches}.append(\Var{targets}) \\
  \Var{allTargets} = \Var{allTargets}[len(\Var{targets}):] \\
}
\Return \Var{batches} \\

\Fn{\FSearch{\Var{targets}, \Var{cutoff}, \Var{modelName}}} {
  \Var{low} = 0, \Var{high} = len(\Var{targets}) - 1, \Var{cutoffIndex} = 0 \\
  \While{low <= high} {
    mid = (low + high) / 2 \\
    estimate = \FEstimate{\Var{targets}[:\Var{mid} + 1], \Var{modelName}} \\
    \If{\Var{estimate} < \Var{cutoff}} {
      \Var{cutoffIndex} = \Var{mid} \\
      \Var{low} = \Var{mid} + 1 \\
    } \lElse {
      \Var{high} = \Var{mid} - 1
    }
  }
  \Return \Var{targets}[:\Var{cutoffIndex} + 1]
}

\end{algorithm}

BTBS generates a set of target batches for each group of
lexicographically sorted targets.
Algorithm~\ref{alg:batching-targets} shows the target batching
algorithm.  The algorithm takes as input a group of sorted targets
\Var{allTargets}, a bound on the max number of targets per batch
\Var{maxTargetsPerBatch}, the memory cutoff value \Var{memoryCutoff}
above which to consider the build with the batch of targets running
out of memory, and the occupancy cutoff value \Var{occupancyCutoff}
above which to consider the build with the batch of targets exceeding
the deadline.  The output \Var{batches} is a list of target batches.

The algorithm first initializes \Var{batches} to an empty list.  When
\Var{allTargets} is not empty, the algorithm takes the first
\Var{maxTargetsPerBatch} targets in \Var{allTargets} as the initial
\Var{targets} list for binary search.  The first binary search queries
the memory model and tries to find the largest sublist of
\Var{targets} that fits into the \Var{memoryCutoff} during execution.
The second binary search queries the occupancy model and tries to find
the largest sublist of \Var{targets} (updated in the first binary
search) that uses less than \Var{occupancyCutoff} ESU.  The binary
searched \Var{targets} list always starts with the same initial target
in each iteration.  Finally, the batch of \Var{targets} is added to
\Var{batches} and removed from \Var{allTargets}.  \XComment{It is
  worth mentioning that Algorithm~\ref{alg:batching-targets} is
  implemented in a memory efficient and multi-threaded way to speedup
  binary searching a large list of targets.}

\CodeIn{limitBatchSizeByCutoff} is the method that does the binary
search.  It takes as input a list of \Var{targets} to search, a
\Var{cutoff} value to fit the final target list and a model name
\Var{modelName} to query the machine learning model.  The
\CodeIn{getEstimateFromTargets} method generates the features from a
synthetic build with the same execution context, build flags as
present in the first \CodeIn{EnqueueTargets} request, but with a
different sublist of \Var{targets}.  Then, the method sends the
generated features to the model which then returns back an estimate.
For the memory model, it returns the estimated peak memory usage of
the build.  For the occupancy model, it returns the average occupancy
of the build.  Finally, the \CodeIn{limitBatchSizeByCutoff} method
returns a sublist of \Var{targets} that uses less than \Var{cutoff} GB
of memory at peak or fewer than \Var{cutoff} ESU on average.  Note
that the returned \Var{targets} sublist always includes at least one
target.

\subsection{Batch Size Reasons}
\label{sec:sec:batch-size-reasons}

\begin{table*}[!t]
  \centering
  \caption{Batch size reasons}
  \label{tab:batch-size-reasons}
  \normalsize
  \begin{tabular}{ >{\centering\arraybackslash}m{5cm} @{\hspace{6ex}}>{\raggedright\arraybackslash}m{12cm} } 
  \hline
  \TTitle{Batch Size Reason} & \multicolumn{1}{c}{\TTitle{Description}} \\
  \hline
  ONLY\_ONE\_TARGET & Remaining unbatched target list only has one target. \\
  MAX\_TARGETS & Initial target batch with size \Var{maxTargetsPerBatch} is valid. \\
  ALL\_REMAINING\_TARGETS & Target batch includes all remaining unbatched targets and is valid. \\
  MAX\_MEMORY & Target batch is created by a binary search with the memory model. \\
  MAX\_OCCUPANCY & Target batch is created by a binary search with the occupancy model. \\
  MEMORY\_ESTIMATE\_ERROR & Target batch includes a fallback size of unbatched targets due to memory error. \\
  OCCUPANCY\_ESTIMATE\_ERROR & Target batch includes a fallback size of unbatched targets due to occupancy error. \\
  \hline
\end{tabular}

\end{table*}

Each target batch is associated with a batch size reason, indicating
why the batch of targets is created.  We define a target batch to be
valid if either (1) it uses less than \Var{memoryCutoff} memory and
\Var{occupancyCutoff} occupancy; or (2) it only has one target.
Table~\ref{tab:batch-size-reasons} shows all possible kinds of batch
size reasons.  \CodeIn{ONLY\_ONE\_TARGET} means that the remaining
unbatched target list only has one target, and BTBS simply creates a
single target batch.  \CodeIn{MAX\_TARGETS} means that the initial
target batch with size \Var{maxTargetsPerBatch} is valid.
\CodeIn{ALL\_REMAINING\_TARGETS} means that the target batch includes
all remaining unbatched targets and it is valid.  \CodeIn{MAX\_MEMORY}
means that the target batch is created by a binary search with the
memory model and is valid.  \CodeIn{MAX\_OCCUPANCY} means that the
target batch is created by a binary search with the occupancy model
and is valid.  \CodeIn{MEMORY\_ESTIMATE\_ERROR} means that the memory
model returns an error and the target batch includes a list of targets
with a fallback size.  \CodeIn{OCCUPANCY\_ESTIMATE\_ERROR} means that
the occupancy model returns an error and the target batch includes a
list of targets with a fallback size.  These batch size reasons can
help us understand the impact of the memory model and occupancy model.

Note that a target batch may have a higher estimate than
\Var{memoryCutoff} or \Var{occupancyCutoff} when it only contains a
single target.  But there is nothing BTBS can do because it cannot
split a single target.  The memory and occupancy models are served in
a separate server and BTBS queries those models via RPC.  So it is
possible that the RPC fails, e.g. RPC deadline exceeded errors.

\subsection{Create Build}
\label{sec:sec:create-build}

For each target batch, BTBS creates a build with the same execution
context and flags as present in the first \CodeIn{EnqueueTargets}
request.  The build will be enqueued in the build service
system~\cite{scalable-build-service}.  Finally, BTBS sends back all
build request IDs generated by the build service system to the clients
so that they can use them to track the build statuses.  The clients
receive a sequence of \CodeIn{EnqueueTargets} responses and each
response contains a single build request ID.

\subsection{Build Failure Retry}
\label{sec:sec:build-failure-retry}

Even with the memory model and the occupancy model, builds may still
fail with OOM or DE errors.  If that happens, BTBS can retry those
builds on the build service system~\cite{scalable-build-service}.  For
an OOM build, BTBS splits the targets in that build into half and
reruns Algorithm~\ref{alg:batching-targets} for each subset of the
split targets.  BTBS does not retry single target OOM builds.  For a
DE build, BTBS reruns Algorithm~\ref{alg:batching-targets} with the
exact same targets in that DE build.  The reason is that we expect the
build service system to cache some actions done for the DE build, so
rerunning the build should be faster and may finish within the
deadline.

\section{Memory and Occupancy Prediction}
\label{sec:memory-occupancy-prediction}

\subsection{Data Collection}
\label{sec:sec:data-collection}

Bazel is a Java based build system and it runs on JVM.  The Bazel
process keeps track of the memory and occupancy usage of a build
execution.

In terms of the memory usage, Bazel records both the peak heap memory
usage and the peak post full GC heap memory usage.  The peak heap
memory usage is the max heap memory usage during the build execution.
Bazel does not necessarily run out of memory if the peak heap memory
usage is close to the allocated memory, because JVM can garbage
collect (GC) the unused references and free some heap memory.  The
peak post full GC heap memory usage is the max heap memory usage after
a full GC during the build execution.  If the peak post full GC heap
memory usage is close to the allocated memory, then Bazel will run out
of memory because GC cannot free up more memory space.  We use the
peak post full GC heap memory usage as the predicting label for the
memory model.  If no GC is triggered during the build execution, then
we use the peak heap memory usage as the predicting label.

In terms of the occupancy usage, Bazel records the exact build
execution time and the total executor service time for the build.  The
build execution time is the wall time of the command-specific
execution, excluding Bazel startup time.  The executor service time
for a Bazel action is the number of executors (or equivalent memory
unified by ESU) multiplied by their execution time occupied by that
action.  The total executor service time for a build is the sum of
executor service time for all actions generated by that build.  The
average executor occupancy is equal to the total executor service time
divided by the build execution time.  Conceptually, the average
executor occupancy measures the average concurrently used executors or
memory during a build execution.  If the average executor occupancy is
above the number of concurrent executors limit, then it is more likely
to cause a \CodeIn{Type I} deadline exceeded build.  We use the
average executor occupancy (in ESU) as the predicting label for the
occupancy model.

The build flags and targets are the raw features we use to predict the
memory and occupancy usage.  The Bazel process is configured to store
all execution statistics to a distributed file
system~\cite{ghemawat2003google} and we have a
Flume~\cite{chambers2010flumejava} job to generate all features and
labels from the execution statistics every day.

\subsection{Feature Engineering}
\label{sec:sec:feature-engineering}

By measuring how important a feature is in predicting the label, we
can choose the most representative features instead of all the
features.  The benefits of feature reduction include (1) speeding up
the model training, (2) making the model training stable (training the
same model multiple times produces close results), (3) avoiding
learning bad features, and (4) reduced data processing, etc.  We use
the mutual information~\cite{kraskov2004estimating,ross2014mutual} to
directly measure how much information a set of features can provide to
predict the label.

\begin{table*}[!t]
  \centering
  \caption{Important features}
  \label{tab:important-features}
  \normalsize
  \begin{tabular}{ >{\raggedright\arraybackslash}m{3.5cm} @{\hspace{6ex}}>{\raggedright\arraybackslash}m{13cm} } 
  \hline
  \multicolumn{1}{c}{\TTitle{Features}} & \multicolumn{1}{c}{\TTitle{Description}} \\
  \hline
  Build priority & The priority of the build. \\
  Command name & The Bazel command name. \\
  Originating user & The username that requests the build. \\
  Product area & The product area under which the build is charged. \\
  Tool tag & The tool that sends targets to BTBS. \\
  Targets & The set of targets in the build. \\
  Target count & The number of distinct targets in the build. \\
  Packages & The set of packages for the corresponding targets in the build. \\
  Package count & The number of distinct packages of all targets in the build. \\
  %% --build\_runfile\_links & Whether the runfiles symlinks for tests and binaries should be in the output directory. \\
  %% --build\_tests\_only & Whether to only build what is necessary to run the test targets. \\
  --cache\_test\_results & Whether Bazel uses the previously saved test results when running tests. \\
  %% --compile\_only & Whether to only run compilation steps related to the targets. \\
  --cpu & The target CPU architecture to be used for the compilation of binaries during the build. \\
  --discard\_analysis\_cache & Whether Bazel should discard the analysis cache right before execution starts. \\
  %% --embed\_label & Include the specified label to the built binary. \\
  %% --flaky\_test\_attempts & The maximum number of times a test should be attempted if it fails. \\
  %% --forge & Whether to run all build and test actions on the remote executor cluster. \\
  --fuseless\_output & Whether to generate intermediate output files in memory. \\
  --jobs & The limit on the number of concurrent executors used during the build execution. \\
  --keep\_going & Whether to proceed as much as possible even in the face of errors. \\
  --keep\_state\_after\_build & Whether to keep incremental in-memory state after the build execution. \\
  %% --local\_fallback & Whether to run build actions locally if Bazel cannot use the remote executor cluster. \\
  --runs\_per\_test & The number of times each test should be executed. \\
  %% --stamp & Whether to include the version information to the built binary. \\
  %% --strip & Whether to strip debugging information from all binaries and libraries. \\
  --test\_size\_filters & Only run test targets with the specified size. \\
  %% --test\_tag\_filters & Only run test targets with the specified tags. \\
  %% --test\_timeout\_filters & Only run test targets with the specified timeout. \\
  --use\_action\_cache & Whether to use Bazel's local action cache. \\
  \hline
\end{tabular}

\end{table*}

Table~\ref{tab:important-features} shows the set of important features
we use to predict both memory and occupancy usages.  The build
priority does not affect the memory or occupancy usage of a build.
But it happens that high priority builds often run a smaller set of
targets compared to low priority builds.  For example, the number of
targets triggered by a human code change is often smaller than the
number of targets triggered by a code coverage tool.  The Bazel
command name is important because it affects the way targets are
built.  For example, the \CodeIn{test} command not only builds the
libraries/binaries but also runs the tests.  As a comparison, the
\CodeIn{build} command only builds a more restricted set, such as
libraries and binaries.  The originating user and the product area are
important because certain users or teams often trigger targets of the
same project and their builds often have similar cost.  The tool tag
is another important feature because the sets of targets triggered by
different tools can vary a lot.  For example, a presubmit service
often triggers a small set of targets given a code change, but the
postsubmit service often triggers millions of targets.  The targets
and the packages, i.e. directories within which the targets are
declared, are very important because targets often have different
costs to build.  Typically, a build with more targets and packages is
more expensive than a build with fewer targets and packages.  Build
flags are also important because they affect how Bazel builds the
targets.  For example, the \CodeIn{--discard\_analysis\_cache} flag
tells Blaze to discard the analysis cache for the previous build
before the next build execution starts, so it affects the memory used
by the build.  The \CodeIn{--keep\_going} flag tells Bazel to proceed
even in the face of errors and it may cause a failed build to use more
memory compared to exiting the execution once a failure occurs.  The
\CodeIn{--runs\_per\_test} flag controls the number of times each test
should be executed so it can affect the number of concurrently used
executors in the case of parallel execution.  The
\CodeIn{--test\_size\_filters} flag filters a subset of test targets
to execute and thus can affect both the memory and occupancy usage of
a build.  In general, the important features selected based on the
mutual information indeed affect the memory and occupancy usage,
intuitively. There are more important flags but we do not describe
them for brevity.

In addition to the basic important features in
Table~\ref{tab:important-features}, we also generate synthetic
features to improve the model accuracy.  For example, the target and
package counts are discretized into quantile buckets.  Specifically,
we can split the range of target counts by their median, and half of
the target counts would fall into the first bucket and the rest half
would fall into the second bucket.  This strategy converts a numeric
feature into a categorical feature.  Another example is to split each
target path into multiple fragments based on the path delimiter.
Specifically, we generate the target prefix splits feature from a
target \CodeIn{//a/b/c:t} by splitting its path into
\CodeIn{//a/b/c:t}, \CodeIn{//a/b/c}, \CodeIn{//a/b} and \CodeIn{//a}.
This helps the model to learn the memory or occupancy usage patterns
for targets under different directories.

All basic and synthetic categorical features are used to generate
feature crosses (Section~\ref{sec:sec:feature-cross}).  We use an
off-the-shelf blackbox optimization tool similar to
AutoML~\cite{hutter2019automated} to generate a set of candidate
feature crosses.  The tool trains the model with a small set of steps
to see which feature crosses give the best performance.  All feature
crosses are generated from the basic and synthetic categorical
features, and we limit the max number of feature crosses to 6.
Finally, the tool returns the best feature crosses for the models.  We
run feature cross search separately for the memory and occupancy
models.

\subsection{Model Training}
\label{sec:sec:model-training}

Once all features are finalized, they will be used to train the memory
and occupancy models.  We choose to train a regression model because
it fits well with the binary search as described in
Algorithm~\ref{alg:batching-targets}.  Another reason to not train a
binary classification model is that the model may perform badly when
only few OOM builds are present in the training data.

We want our models to be monotonic over all targets related features.
This property makes sure that adding new targets to the build with the
same flags always results in an increased memory estimate.  If the
monotonicity property does not hold, then the binary search may not
work as expected.  We achieve the monotonicity property by setting the
regularization term to infinity when the weights of the targets
related features are negative.  Specifically, we use the loss function
as follows:

{
\setlength{\abovedisplayskip}{0pt}
\setlength{\belowdisplayskip}{0pt}
\begin{align*}
L(\boldsymbol{\beta},\mathbf{x},\mathbf{y}) &=
\frac{1}{n}\sum_{i=1}^{n}(y_{i}-\mathbf{x}_{i}^{\mathsf{T}}{\boldsymbol{\beta}})
+ R(\boldsymbol{\beta}) \\
R(\boldsymbol{\beta}) &= \sum_{i=1}^{n}r_{i}(\boldsymbol{\beta_{i}}) \\
r_{i}(\boldsymbol{\beta_{i}}) &= \begin{cases}
\boldsymbol{\lambda_{i}^{-}}|\boldsymbol{\beta_{i}}| & \boldsymbol{\beta_{i}} < 0 \\
\boldsymbol{\lambda_{i}^{+}}|\boldsymbol{\beta_{i}}| & \boldsymbol{\beta_{i}} \geq 0
\end{cases} \\
\boldsymbol{\lambda_{i}^{-}}, \boldsymbol{\lambda_{i}^{+}} &\geq 0
\end{align*}
}

\noindent{}where $L(\boldsymbol{\beta},\mathbf{x},\mathbf{y})$ is the
mean squared error (MSE) loss function and $R(\boldsymbol{\beta})$ is
the regularization term.  $\boldsymbol{\lambda_{i}^{-}}$ and
$\boldsymbol{\lambda_{i}^{+}}$ are the L1 regularization term for the
weight $\boldsymbol{\beta_{i}}$ when $\boldsymbol{\beta_{i}} < 0$ and
$\boldsymbol{\beta_{i}} \geq 0$, respectively.  We set
$\boldsymbol{\lambda_{i}^{-}} = \infty$ for the weights of the targets
related features.  This causes all the weights of the targets related
features to be non-negative, thus making the model monotonic over all
targets related features.

We train the memory and occupancy models using the last 17 days of all
builds that are executed in the build service system.  This means that
the training and testing datasets include builds that are not created
by BTBS.  17 is chosen to cover at least 2 weeks of weekday data and
be able to handle holiday weekends.  However, using a long period of
training data may cause the memory model to slowly capture the new
memory patterns of recent builds.  For example, it is possible that a
memory usage regression is injected into Bazel, which causes otherwise
identical builds to use more memory than before.  The memory model
won't be able to capture the memory usage increase until a couple of
days later because a majority of the builds still use less memory.  To
solve this problem, we train another memory model that uses data from
the most recent day and BTBS uses the maximum memory estimate of the
two models when performing the binary search.  All the models are
continuously trained and pushed to production.

\section{Evaluation}
\label{sec:evaluation}

\XComment{In this section, we first discuss the production setup of
  BTBS (Section~\ref{sec:sec:production-setup}).  Then, we discuss the
  performance of BTBS (Section~\ref{sec:sec:performance}).  Finally,
  we discuss the model impact (Section~\ref{sec:sec:model-impact}) and
  accuracy (Section~\ref{sec:sec:model-accuracy}).}

\subsection{Production Setup}
\label{sec:sec:production-setup}

BTBS is deployed geographically at 3 locations in the US and each
location has 5 running jobs that serve traffic all over the world.
This distributed setting avoids single point failure.  Our experiment
lasts from 2020/01/16 to 2020/02/19 (35 days).  We report the
performance of BTBS using the production data.  During the time, BTBS
receives 51 million \CodeIn{EnqueueTargets} RPC calls (1.47 million
daily on average) and creates 102 million builds (2.92 million daily
on average).

The \Var{maxTargetsPerBatch} is set to 900 in
Algorithm~\ref{alg:batching-targets}.  The memory cutoff value is set
to 7GB for high priority builds, 9GB for medium priority builds and
10GB for low priority builds.  The reason is that the memory model is
not 100\% accurate and the number of OOM builds will increase if we
set the cutoff close to the allocated memory (13GB).  We set a lower
memory cutoff for high priority builds because we cannot afford the
human users to wait for OOM failure retries.  In contrast, we can
tolerate more low priority OOM builds, e.g. builds that collect code
coverage and do not block developers.  The occupancy cutoff value is
set to 500 ESU which is smaller than the max allowed concurrent
executors limit (600).  If the memory or occupancy model returns
errors, then we fallback to a default batch size of 300.

All regularization terms $\boldsymbol{\lambda_{i}^{+}}, i=1...n$ for
positive weights are set to 7000.  All the learning parameters,
e.g. learning rate and batch size, are set to the default values.
During the experiment time, the build service system ran 17.8 millions
builds on average each day and 34.8\% builds were Bazel queries that
do not use much memory or any executor.  So we excluded those builds
and used the rest 11.6 million builds as the training examples daily.

In the rest of this section, 1k represents 1 thousand and 1m
represents 1 million.  The error is calculated as the difference
between the actual and estimated memory usage, i.e. $(y-\hat{y})$.

\subsection{BTBS Performance}
\label{sec:sec:performance}

Table~\ref{tab:performance} shows some metrics to measure the
performance of BTBS.  \emph{QPS} and \emph{Latency} show the queries
per second and latency in milliseconds of the \CodeIn{EnqueueTargets}
streaming RPC over the past 24 hours during the entire experiment
span.  \emph{StreamTC} shows the total target count per RPC to BTBS.
\emph{StreamBC} shows the total generated build count per RPC to BTBS.
\emph{BuildTC}, \emph{ExecT}, \emph{Memory} and \emph{Occupancy} show
the target count, execution time in seconds, memory usage in gigabytes
and occupancy usage in ESU of individual builds created by BTBS.

BTBS is used heavily in production and the average QPS over the past
24 hours ranges from \UseMacro{qps-avg-min} to \UseMacro{qps-avg-max}.
The min and max QPS over the past 24 hours range from
\UseMacro{qps-min-min} to \UseMacro{qps-min-max} and from
\UseMacro{qps-max-min} to \UseMacro{qps-max-max}, respectively.  BTBS
receives less traffic on weekends off the peak hours and more traffic
on weekdays during the peak hours.  The batching algorithm is
efficient and the average latency over the past 24 hours ranges from
\UseMacro{latency-avg-min} to \UseMacro{latency-avg-max} milliseconds.
The min and max latency over the past 24 hours range from
\UseMacro{latency-min-min} to \UseMacro{latency-min-max} milliseconds
and from \UseMacro{latency-max-min} to \UseMacro{latency-max-max}
milliseconds, respectively.  The latency of BTBS is very small if the
clients enqueue less than hundreds of targets, but it could go up to
minutes if the clients enqueue millions of targets.  The average, min,
median and max number of targets received by BTBS per RPC are
\UseMacro{stream-target-count-avg},
\UseMacro{stream-target-count-min},
\UseMacro{stream-target-count-median} and
\UseMacro{stream-target-count-max}, respectively.  A majority of the
\CodeIn{EnqueueTargets} requests only contain 1-2 targets and they are
sent by tools like presubmit failure rerunner, culprit finder, flaky
test detector, etc.  The postsubmit service could send millions of
targets to BTBS because it needs to make sure that all tests can pass
at a given revision.  The average, min, median and max number of
builds created by BTBS per RPC are \UseMacro{stream-build-count-avg},
\UseMacro{stream-build-count-min},
\UseMacro{stream-build-count-median} and
\UseMacro{stream-build-count-max}, respectively.  This is consistent
with the number of targets received by BTBS because BTBS creates more
builds as it receives more targets per RPC.  Since a majority of the
\CodeIn{EnqueueTargets} requests only contain 1-2 targets, BTBS only
needs to create a single build for most of the time.  The average,
min, median and max number of targets per build created by BTBS are
\UseMacro{build-target-count-avg}, \UseMacro{build-target-count-min},
\UseMacro{build-target-count-median} and
\UseMacro{build-target-count-max}, respectively.  Note that the number
of targets per build is bounded by the \Var{maxTargetsPerBatch}.  The
average, min, median and max execution time of builds created by BTBS
are \UseMacro{build-execution-time-avg},
\UseMacro{build-execution-time-min},
\UseMacro{build-execution-time-median} and
\UseMacro{build-execution-time-max} seconds, respectively.  It
indicates that most builds run with a couple of seconds to minutes but
some builds can run with 1-2 hours.  A majority of builds have a
deadline of 1.5 hours and will be expired if they do not finish by the
deadline.  The average, min, median and max memory usage of builds
created by BTBS are \UseMacro{build-memory-avg},
\UseMacro{build-memory-min}, \UseMacro{build-memory-median} and
\UseMacro{build-memory-max} GB, respectively.  It shows that many
builds take less than a few GB of memory but some of them can still
use up all the allocated memory.  The average, min, median and max
occupancy of builds created by BTBS are
\UseMacro{build-occupancy-avg}, \UseMacro{build-occupancy-min},
\UseMacro{build-occupancy-median} and \UseMacro{build-occupancy-max}
ESU, respectively.  The max occupancy usage is greater than the max
allowed concurrent executors because some builds use more executor
memory.  It shows that a majority of builds only use a few concurrent
executors or executor memory but some builds can use thousands ESU.
It is worth mentioning that some builds hit the cached actions so they
have 0 occupancy usage.

\XComment{In practice, the extreme variance in memory and occupancy
  usage of targets motivates us to take a machine learning approach.
  If all targets have similar memory and occupancy usage, we could
  simply put \Var{maxTargetsPerBatch} targets in a build and be
  largely done.}

\begin{table}[!t]
  \centering
  \caption{BTBS performance}
  \label{tab:performance}
  \normalsize
  \begin{tabular}{C{19mm}R{14mm}R{14mm}R{14mm}R{16mm}}
  \hline
  \TTitle{Metrics} & \TTitle{Avg} & \TTitle{Min} & \TTitle{Median} & \TTitle{Max} \\
  \hline
  QPS & \UseMacro{qps-avg-min}-\UseMacro{qps-avg-max} & \UseMacro{qps-min-min}-\UseMacro{qps-min-max} & \UseMacro{qps-median-min}-\UseMacro{qps-median-max} & \UseMacro{qps-max-min}-\UseMacro{qps-max-max} \\
  Latency & \UseMacro{latency-avg-min}-\UseMacro{latency-avg-max} & \UseMacro{latency-min-min}-\UseMacro{latency-min-max} & \UseMacro{latency-median-min}-\UseMacro{latency-median-max} & \UseMacro{latency-max-min}-\UseMacro{latency-max-max} \\
  StreamTC & \UseMacro{stream-target-count-avg} & \UseMacro{stream-target-count-min} & \UseMacro{stream-target-count-median} & \UseMacro{stream-target-count-max} \\
  StreamBC & \UseMacro{stream-build-count-avg} & \UseMacro{stream-build-count-min} & \UseMacro{stream-build-count-median} & \UseMacro{stream-build-count-max} \\
  BuildTC & \UseMacro{build-target-count-avg} & \UseMacro{build-target-count-min} & \UseMacro{build-target-count-median} & \UseMacro{build-target-count-max} \\
  ExecT & \UseMacro{build-execution-time-avg} & \UseMacro{build-execution-time-min} & \UseMacro{build-execution-time-median} & \UseMacro{build-execution-time-max} \\
  Memory & \UseMacro{build-memory-avg} & \UseMacro{build-memory-min} & \UseMacro{build-memory-median} & \UseMacro{build-memory-max} \\
  Occupancy & \UseMacro{build-occupancy-avg} & \UseMacro{build-occupancy-min} & \UseMacro{build-occupancy-median} & \UseMacro{build-occupancy-max} \\
  \hline
\end{tabular}

\end{table}

\subsection{Model Impact}
\label{sec:sec:model-impact}

\begin{table*}[!t]
  \centering
  \caption{Build count distribution by batch size reasons}
  \label{tab:build-count-distribution}
  \normalsize
  \begin{tabular}{C{5cm}R{1.5cm}R{2cm}R{2cm}R{1.5cm}R{1.5cm}R{1.5cm}}
  \hline
  \TTitle{Batch Size Reason} & \TTitle{\#Build} & \TTitle{\#OOM(\%)} & \TTitle{\#DE(\%)} & \TTitle{\%Build} & \TTitle{\%OOM} & \TTitle{\%DE} \\
  \hline
  ONLY\_ONE\_TARGET & \UseMacro{only-one-target-build-cnt} & \UseMacro{only-one-target-oom-cnt} (\UseMacro{only-one-target-oom-cnt-ratio}) & \UseMacro{only-one-target-de-cnt} (\UseMacro{only-one-target-de-cnt-ratio}) & \UseMacro{only-one-target-build-percentage} & \UseMacro{only-one-target-oom-percentage} & \UseMacro{only-one-target-de-percentage} \\
  MAX\_TARGETS & \UseMacro{max-targets-build-cnt} & \UseMacro{max-targets-oom-cnt} (\UseMacro{max-targets-oom-cnt-ratio}) & \UseMacro{max-targets-de-cnt} (\UseMacro{max-targets-de-cnt-ratio}) & \UseMacro{max-targets-build-percentage} & \UseMacro{max-targets-oom-percentage} & \UseMacro{max-targets-de-percentage} \\
  ALL\_REMAINING\_TARGETS & \UseMacro{all-remaining-targets-build-cnt} & \UseMacro{all-remaining-targets-oom-cnt} (\UseMacro{all-remaining-targets-oom-cnt-ratio}) & \UseMacro{all-remaining-targets-de-cnt} (\UseMacro{all-remaining-targets-de-cnt-ratio}) & \UseMacro{all-remaining-targets-build-percentage} & \UseMacro{all-remaining-targets-oom-percentage} & \UseMacro{all-remaining-targets-de-percentage} \\
  MAX\_MEMORY & \UseMacro{max-memory-build-cnt} & \UseMacro{max-memory-oom-cnt} (\UseMacro{max-memory-oom-cnt-ratio}) & \UseMacro{max-memory-de-cnt} (\UseMacro{max-memory-de-cnt-ratio}) & \UseMacro{max-memory-build-percentage} & \UseMacro{max-memory-oom-percentage} & \UseMacro{max-memory-de-percentage} \\
  MAX\_OCCUPANCY & \UseMacro{max-occupancy-build-cnt} & \UseMacro{max-occupancy-oom-cnt} (\UseMacro{max-occupancy-oom-cnt-ratio}) & \UseMacro{max-occupancy-de-cnt} (\UseMacro{max-occupancy-de-cnt-ratio}) & \UseMacro{max-occupancy-build-percentage} & \UseMacro{max-occupancy-oom-percentage} & \UseMacro{max-occupancy-de-percentage} \\
  MEMORY\_ESTIMATE\_ERROR & \UseMacro{memory-estimate-error-build-cnt} & \UseMacro{memory-estimate-error-oom-cnt} (\UseMacro{memory-estimate-error-oom-cnt-ratio}) & \UseMacro{memory-estimate-error-de-cnt} (\UseMacro{memory-estimate-error-de-cnt-ratio}) & \UseMacro{memory-estimate-error-build-percentage} & \UseMacro{memory-estimate-error-oom-percentage} & \UseMacro{memory-estimate-error-de-percentage} \\
  OCCUPANCY\_ESTIMATE\_ERROR & \UseMacro{occupancy-estimate-error-build-cnt} & \UseMacro{occupancy-estimate-error-oom-cnt} (\UseMacro{occupancy-estimate-error-oom-cnt-ratio}) & \UseMacro{occupancy-estimate-error-de-cnt} (\UseMacro{occupancy-estimate-error-de-cnt-ratio}) & \UseMacro{occupancy-estimate-error-build-percentage} & \UseMacro{occupancy-estimate-error-oom-percentage} & \UseMacro{occupancy-estimate-error-de-percentage} \\
  \hline
  \TTitle{Sum} & \UseMacro{sum-build-cnt} & \UseMacro{sum-oom-cnt} (\UseMacro{sum-oom-cnt-ratio}) & \UseMacro{sum-de-cnt} (\UseMacro{sum-de-cnt-ratio}) & 100 & 100 & 100 \\
  \hline
\end{tabular}

\end{table*}

\begin{table*}[!t]
  \centering
  \caption{Build stats by batch size reasons}
  \label{tab:build-stats}
  \normalsize
  \begin{tabular}{C{5cm}R{1.5cm}R{1.2cm}R{1.5cm}R{1.5cm}|R{1.5cm}R{1.2cm}R{1.5cm}R{1.5cm}}
  \hline
  \multirow{2}{*}{\TTitle{Batch Size Reason}} & \multicolumn{4}{c|}{\TTitle{Avg}} & \multicolumn{4}{c}{\TTitle{Median}} \\
  & \TTitle{BuildTC} & \TTitle{ExecT} & \TTitle{Memory} & \TTitle{Occup} & \TTitle{BuildTC} & \TTitle{ExecT} & \TTitle{Memory} & \TTitle{Occup} \\
  \hline
  ONLY\_ONE\_TARGET & \UseMacro{only-one-target-avg-target-cnt} & \UseMacro{only-one-target-avg-execution-time} & \UseMacro{only-one-target-avg-memory} & \UseMacro{only-one-target-avg-occupancy} & \UseMacro{only-one-target-median-target-cnt} & \UseMacro{only-one-target-median-execution-time} & \UseMacro{only-one-target-median-memory} & \UseMacro{only-one-target-median-occupancy} \\
  MAX\_TARGETS & \UseMacro{max-targets-avg-target-cnt} & \UseMacro{max-targets-avg-execution-time} & \UseMacro{max-targets-avg-memory} & \UseMacro{max-targets-avg-occupancy} & \UseMacro{max-targets-median-target-cnt} & \UseMacro{max-targets-median-execution-time} & \UseMacro{max-targets-median-memory} & \UseMacro{max-targets-median-occupancy} \\
  ALL\_REMAINING\_TARGETS & \UseMacro{all-remaining-targets-avg-target-cnt} & \UseMacro{all-remaining-targets-avg-execution-time} & \UseMacro{all-remaining-targets-avg-memory} & \UseMacro{all-remaining-targets-avg-occupancy} & \UseMacro{all-remaining-targets-median-target-cnt} & \UseMacro{all-remaining-targets-median-execution-time} & \UseMacro{all-remaining-targets-median-memory} & \UseMacro{all-remaining-targets-median-occupancy} \\
  MAX\_MEMORY & \UseMacro{max-memory-avg-target-cnt} & \UseMacro{max-memory-avg-execution-time} & \UseMacro{max-memory-avg-memory} & \UseMacro{max-memory-avg-occupancy} & \UseMacro{max-memory-median-target-cnt} & \UseMacro{max-memory-median-execution-time} & \UseMacro{max-memory-median-memory} & \UseMacro{max-memory-median-occupancy} \\
  MAX\_OCCUPANCY & \UseMacro{max-occupancy-avg-target-cnt} & \UseMacro{max-occupancy-avg-execution-time} & \UseMacro{max-occupancy-avg-memory} & \UseMacro{max-occupancy-avg-occupancy} & \UseMacro{max-occupancy-median-target-cnt} & \UseMacro{max-occupancy-median-execution-time} & \UseMacro{max-occupancy-median-memory} & \UseMacro{max-occupancy-median-occupancy} \\
  MEMORY\_ESTIMATE\_ERROR & \UseMacro{memory-estimate-error-avg-target-cnt} & \UseMacro{memory-estimate-error-avg-execution-time} & \UseMacro{memory-estimate-error-avg-memory} & \UseMacro{memory-estimate-error-avg-occupancy} & \UseMacro{memory-estimate-error-median-target-cnt} & \UseMacro{memory-estimate-error-median-execution-time} & \UseMacro{memory-estimate-error-median-memory} & \UseMacro{memory-estimate-error-median-occupancy} \\
  OCCUPANCY\_ESTIMATE\_ERROR & \UseMacro{occupancy-estimate-error-avg-target-cnt} & \UseMacro{occupancy-estimate-error-avg-execution-time} & \UseMacro{occupancy-estimate-error-avg-memory} & \UseMacro{occupancy-estimate-error-avg-occupancy} & \UseMacro{occupancy-estimate-error-median-target-cnt} & \UseMacro{occupancy-estimate-error-median-execution-time} & \UseMacro{occupancy-estimate-error-median-memory} & \UseMacro{occupancy-estimate-error-median-occupancy} \\
  \hline
\end{tabular}

\end{table*}

Table~\ref{tab:build-count-distribution} shows the build count
distribution by batch size reasons.  \emph{\#Build} shows the number
of builds.  \emph{\#OOM(\%)} shows the number of OOM builds and its
percentage out of all builds in each batch size reason.
\emph{\#DE(\%)} shows the number of DE builds and its percentage out
of all builds in each batch size reason.  \emph{\%Build} shows the
percentage of builds in each batch size reason out of all builds.
\emph{\%OOM} shows the percentage of OOM builds in each batch size
reason out of all OOM builds.  \emph{\%DE} shows the percentage of DE
builds in each batch size reason out of all DE builds.

Builds with batch size reasons \CodeIn{ONLY\_ONE\_TARGET},
\CodeIn{MAX\_TARGETS}, \CodeIn{ALL\_REMAINING\_TARGETS},
\CodeIn{MAX\_MEMORY}, \CodeIn{MAX\_OCCUPANCY},
\CodeIn{MEMORY\_ESTIMATE\_ERROR} and
\CodeIn{OCCUPANCY\_ESTIMATE\_ERROR} account for
\UseMacro{only-one-target-build-percentage}\%,
\UseMacro{max-targets-build-percentage}\%,
\UseMacro{all-remaining-targets-build-percentage}\%,
\UseMacro{max-memory-build-percentage}\%,
\UseMacro{max-occupancy-build-percentage}\%,
\UseMacro{memory-estimate-error-build-percentage}\% and
\UseMacro{occupancy-estimate-error-build-percentage}\% of the total
builds, respectively.  The memory model is used in all builds with
batch size reasons \CodeIn{MAX\_TARGETS},
\CodeIn{ALL\_REMAINING\_TARGETS}, \CodeIn{MAX\_MEMORY},
\CodeIn{MAX\_OCCUPANCY} and \CodeIn{OCCUPANCY\_ESTIMATE\_ERROR}, which
is 78.78\% of all builds.  The occupancy model is used in all builds
with batch size reasons \CodeIn{MAX\_TARGETS},
\CodeIn{ALL\_REMAINING\_TARGETS}, \CodeIn{MAX\_MEMORY},
\CodeIn{MAX\_OCCUPANCY}, which is 78.78\% of all builds.
\CodeIn{ONLY\_ONE\_TARGET} builds use neither the memory model nor the
occupancy model because BTBS always creates a build if a single target
is left in the remaining target list.
\CodeIn{MEMORY\_ESTIMATE\_ERROR} and
\CodeIn{OCCUPANCY\_ESTIMATE\_ERROR} builds may use the memory and
occupancy models, respectively, in the initial binary search
iterations before the failure.  So both the memory and occupancy
models are heavily used in BTBS.

\CodeIn{MAX\_MEMORY} builds provide an indicator on the potential
number of OOM builds if BTBS simply creates builds with
\Var{maxTargetsPerBatch} targets.  It shows that
\UseMacro{max-memory-oom-cnt-ratio}\% of \CodeIn{MAX\_MEMORY} builds
still run out of memory and they account for
\UseMacro{max-memory-oom-percentage}\% of all out of memory builds.
This is expected because the \CodeIn{MAX\_MEMORY} builds should use
more memory than other builds.  The \CodeIn{MEMORY\_ESTIMATE\_ERROR}
builds approximately indicate how BTBS would behave without the memory
model.  It shows that \UseMacro{memory-estimate-error-oom-cnt-ratio}\%
\CodeIn{MEMORY\_ESTIMATE\_ERROR} builds run out of memory.  If the
\CodeIn{MAX\_MEMORY} builds have the same out of memory rate as the
\CodeIn{MEMORY\_ESTIMATE\_ERROR} builds, we would have 74k more out of
memory builds during the experiment.  In practice, the memory model is
very important because an OOM build is very expensive to retry and it
blocks the developer's productivity.  The memory model is useful in
limiting the memory usage while maximizing the batch density.  BTBS
used a small \Var{maxTargetsPerBatch} before the memory model existed,
the problem was that it generated too many builds and used up almost
all Bazel workers, which blocked other builds from running.

\CodeIn{MAX\_OCCUPANCY} builds provide an indicator on the potential
number of \CodeIn{Type I} DE builds if BTBS simply creates builds with
\Var{maxTargetsPerBatch} targets.  It shows that
\UseMacro{max-occupancy-de-cnt-ratio}\% of \CodeIn{MAX\_OCCUPANCY}
builds still exceed the deadlines and they account for
\UseMacro{max-occupancy-de-percentage}\% of all deadline exceeded
builds.  Interestingly, a majority of DE builds are not
\CodeIn{MAX\_OCCUPANCY} builds.  This indicates that most DE builds
are \CodeIn{Type II} DE builds.  Out of all DE builds,
\UseMacro{type-ii-de-percentage}\% builds use less than or equal to
600 ESU and only \UseMacro{type-i-de-percentage}\% builds use more
than 600 ESU.  This confirms that \CodeIn{Type II} DE builds dominate
in the current setup.  Unfortunately, we only have
\UseMacro{occupancy-estimate-error-build-cnt}
\CodeIn{OCCUPANCY\_ESTIMATE\_ERROR} builds and no one exceeds the
deadline, so it's hard to quantitatively measure how many more
deadline exceeded builds we would have without the occupancy model.
In practice, we used to have more DE builds before the occupancy model
exists, so we believe that the occupancy model is useful to reduce
\CodeIn{Type I} DE errors.

\begin{figure*}[!t]
  \centering
  \input{figures/memory-model-accuracy}
  \caption{Memory model accuracy (GB)}
  \label{fig:memory-model-accuracy}
  \vspace{-1em}
\end{figure*}

\begin{figure*}[!t]
  \centering
  \input{figures/occupancy-model-accuracy}
  \caption{Occupancy model accuracy (ESU)}
  \label{fig:occupancy-model-accuracy}
  \vspace{-1em}
\end{figure*}

Table~\ref{tab:build-stats} shows the build target count
(\emph{BuildTC}), execution time in seconds (\emph{ExecT}), memory
usage in GB (\emph{Memory}) and occupancy usage in ESU (\emph{Occup})
on average and median, respectively, for each batch size reason.

The \CodeIn{MAX\_MEMORY} builds use \UseMacro{max-memory-avg-memory}GB
and \UseMacro{max-memory-median-memory}GB memory on average and
median, respectively.  This indicates that \CodeIn{MAX\_MEMORY} builds
use more memory than other builds and the usage is close to the memory
cutoff.  The \CodeIn{MAX\_OCCUPANCY} builds use
\UseMacro{max-occupancy-avg-occupancy} and
\UseMacro{max-occupancy-median-occupancy} occupancy on average and
median, respectively.  This indicates that \CodeIn{MAX\_OCCUPANCY}
builds use more occupancy than other builds and the usage is close to
the occupancy cutoff.  Moreover, it takes
\UseMacro{max-occupancy-avg-execution-time}s and
\UseMacro{max-occupancy-median-execution-time}s to run
\CodeIn{MAX\_OCCUPANCY} builds on average and median, respectively.
This indicates that \CodeIn{MAX\_OCCUPANCY} builds are longer running
than other builds and the execution time is proportional to the
occupancy usage in general.  The result shows that the memory and
occupancy models can restrict the memory and occupancy usage of the
builds, thus reducing the OOM or DE errors.

\CodeIn{ONLY\_ONE\_TARGET} builds are single target builds, so they
typically have the least execution time and memory/occupancy usage as
expected.  All \CodeIn{MAX\_TARGETS} builds have
\UseMacro{max-targets-avg-target-cnt} targets and BTBS cannot generate
builds with more than \Var{maxTargetsPerBatch} targets.  So
\CodeIn{MAX\_TARGETS} builds typically have more execution time and
memory/occupancy usage.  \CodeIn{ALL\_REMAINING\_TARGETS} builds are
generated at the end of Algorithm~\ref{alg:batching-targets}, so these
builds typically have few targets and use little memory and occupancy.
The average build target count for the
\CodeIn{MEMORY\_ESTIMATE\_ERROR} and
\CodeIn{OCCUPANCY\_ESTIMATE\_ERROR} builds is less than 300 because
the initial binary search iterations may already cut the batch size to
be less than 300.  Both \CodeIn{MEMORY\_ESTIMATE\_ERROR} and
\CodeIn{OCCUPANCY\_ESTIMATE\_ERROR} builds are rare and they use
little memory and occupancy.

\subsection{Model Accuracy}
\label{sec:sec:model-accuracy}

Figure~\ref{fig:rmse-by-actual-memory} shows the root mean square
error (RMSE) of the memory model grouped by different actual memory
usages (0GB to 13GB with a step of 0.1GB).  The graph is broken down
by whether the memory model overestimates (blue line), underestimates
(red line) or accurately predicts (yellow line) the actual memory
usage, respectively.  It also shows the overall RMSE (green line).  We
can see that the memory model heavily overestimates builds which use
very little memory.  A common reason is that some targets may finish
without GC so their memory usage is recorded much larger than that
after the GC.  The memory model learns from builds without GC that
some targets have large memory usage, so it often overestimates the
memory usage of builds with the same targets that finish after GC.
The model also overestimates builds which use 7-8GB memory.  Many of
these builds are single target builds and the model gives large memory
estimates for all of these builds.  In general, the model tends to
have a larger error in overestimation compared to underestimation.
The error tends to go up for larger memory builds.
Figure~\ref{fig:build-count-by-memory-error} shows the number of
builds grouped by different error values.  The graph is broken down by
the build priority.  It shows that most builds, regardless of their
priorities, have an error close to 0, which means that the model
performs well.  Figure~\ref{fig:build-count-by-actual-memory} shows
the number of \CodeIn{MAX\_MEMORY} builds grouped by their actual
memory usages.  We can see that the memory usage of high, medium and
low priority builds peaks at around 7GB, 9GB and 10GB, respectively.
This is consistent with our setup.

Figure~\ref{fig:rmse-by-actual-occupancy} shows the RMSE of the
occupancy model grouped by different actual occupancy usages (0 ESU to
500 ESU with a step of 1 ESU).  It seems that the occupancy error
increases for larger occupancy builds.  The overestimation error
increases slowly and then decreases until it diminishes at 500.  The
model only underestimates builds that use more than 500 ESU (not shown
in the graph).  The underestimation error increases for larger
occupancy builds.  The overall RMSE increases as the actual build
occupancy usage increases.  This is expected because most builds use
<100 ESU and the model does not have much data to learn larger
occupancy builds.  Figure~\ref{fig:build-count-by-occupancy-error}
shows the number of builds grouped by different occupancy error
values.  It shows that most builds have an error close to 0, which
shows that the model performs well.
Figure~\ref{fig:build-count-by-actual-occupancy} shows the number of
\CodeIn{MAX\_OCCUPANCY} builds grouped by their actual occupancy
usages.  We can see that the occupancy usage peaks at both 0 and
around 500 ESU.  The peak at 0 ESU is caused by many builds hitting
the action cache in the executor cluster~\cite{scalable-build-service}
and these builds do not occupy any executor.  The peak at around 500
ESU is consistent with our setup.

\section{Related Work}
\label{sec:relatedwork}

We do not find any related work about using machine learning models to
construct builds that use limited resources to reduce OOM and DE
errors.  This problem is quite common in \Company{} and we expect it
to be common in other companies that have a monolithic code repository
and many build targets.  Our experience is that a service like BTBS
becomes more important as the need to build a large number of targets
increases.  We believe that this work provides an insight for
developers who have a similar need, e.g. Bazel~\cite{bazel} and
Buck~\cite{buck} users.

\Company's TAP service~\cite{memon2017taming} mentions that the test
executions could fail due to OOM errors, but it does not give a
solution.  BTBS is designed exactly to solve that problem.  In fact,
the TAP service is one of the BTBS users.

The build service system~\cite{scalable-build-service} describes how
the build is executed remotely in \Company.  BTBS is one of the build
service system users.  It is designed to workaround the constraints,
i.e. limited memory allocation and max allowed concurrent executors,
imposed by the build service system which is not capable of building a
very large number of targets.

\section{Conclusions}
\label{sec:conclusion}

In this paper, we describe the first build target batching service
(BTBS) that is able to partition a large stream of targets into a set
of target batches such that the builds created from those target
batches do not have OOM or DE errors.  We discuss the batching
algorithm as well as the machine learning models used in BTBS.
Overall, we show that BTBS is able to achieve a low OOM rate
(\UseMacro{sum-oom-cnt-ratio}\%) and a low DE rate
(\UseMacro{sum-de-cnt-ratio}\%).  We believe that BTBS introduces
useful insights and can help in designing new target batching systems.

\bibliographystyle{abbrv}
\bibliography{bib}

\end{document}